%% file: main.tex
\documentclass[11pt,a4paper,twoside]{article}

\usepackage[utf8]{inputenc}
\usepackage[all]{nowidow}
\usepackage[font=footnotesize]{caption}

\usepackage[normalem]{ulem}

\usepackage{etoolbox}

\usepackage{a4}
\usepackage[english]{babel}
\usepackage{csquotes}
\usepackage{fancybox,fancyhdr,enumerate}
\usepackage{graphicx}
\usepackage{wrapfig}
\usepackage{float}
\usepackage{lscape}
\usepackage{paralist}
\usepackage{enumitem}
\usepackage{subfig}
\usepackage{titlesec}
\usepackage{tabularx}
\usepackage{longtable}

\usepackage[backend=bibtex,giveninits=true,style=numeric,sorting=none,defernumbers=true,maxnames=1]{biblatex}
\newtoggle{bbx:allnames}
\toggletrue{bbx:allnames}
\AtEveryBibitem{\iftoggle{bbx:allnames}{\defcounter{maxnames}{999}}{}}

\newif\ifsubm
\submfalse  
\newif\ifkurz
\kurztrue

\usepackage[textsize=tiny]{todonotes}

\usepackage{verbatim}

\usepackage{eurosym}
\usepackage{url}
\usepackage{color, soul}
\usepackage{lpic}
\usepackage{amsmath}
\usepackage{amsfonts}

\usepackage{multirow}
\usepackage{multicol}

\usepackage{wrapfig, siunitx}

\usepackage{mdframed}

\usepackage[printonlyused]{acronym}
\input{_acronyms}

\newif\ifsqueeze
\squeezefalse 	
\ifsqueeze
\fi

\usepackage{paralist}
\newenvironment{shortitemize}{\begin{compactitem}}{\end{compactitem}}

\newcommand{\up}{\vspace*{-0.5em}}
\newcommand{\dn}{\vspace*{0.5em}}

\usepackage{helvet}

\begin{document}
\sffamily
\sloppy

\selectlanguage{english}

\pagestyle{fancy}

\lhead[\fancyplain{}{\thepage}]%
             {\fancyplain{}{Towards Organic Distribution Systems}}
\rhead[\fancyplain{}{Towards Organic Distribution Systems}]%
             {\fancyplain{}{\thepage}}
\cfoot{}

\titlespacing*{\subsubsection}{0pt}{2ex}{.5ex} 
\titlespacing*{\subsection}{0pt}{2ex}{.5ex} 
\titlespacing*{\section}{0pt}{3ex}{.5ex} 


\dn\dn\dn

{\Large\bf
\begin{center}
Towards Organic Distribution Systems -- The Vision of Self-Configuring, Self-Organising, Self-Healing, and Self-Optimising Power Distribution Management
\end{center}
}

\dn\dn\dn
{\bf
\begin{center}
 Inga Loeser$^1$, Martin Braun$^{1, 2}$, Christian Gruhl$^3$, Jan-Hendrik Menke$^{1,4}$, Bernhard Sick$^3$, Sven Tomforde$^5$
\end{center}
}
\begin{center}
    $^1$ Energy Management and Power System Operation, University of Kassel, Germany\\
    $^2$ Grid Planning and Operation, Fraunhofer IEE, Germany\\
    $^3$ Intelligent Embedded Systems, University of Kassel, Germany\\
    $^4$ System Development, Amprion GmbH, Germany\\
    $^5$ Intelligent Systems, Christian-Albrechts-Universität zu Kiel, Germany\\
\end{center}

\dn\dn\dn
{\small
\textbf{Abstract.} Due to the decarbonisation of energy use, the power system is expected to become the backbone of all energy sectors and thus the basic critical infrastructure. High penetration with distributed energy resources demands the coordination of a large number of prosumers, partly controlled by home energy management systems (HEMS), to be designed in such a way that the power system's operational limits are not violated. On the grid level, distribution management systems (DMS) try to keep the power system in the normal operational state. On the prosumer level, distributed HEMS optimise the internal power flows by using batteries, photovoltaic generators, or flexible loads optimally. The vision of the ODiS (Organic Distribution System) initiative is to develop an architecture to operate a distribution grid reliably, with high resiliency, and fully autonomously by developing ``organic'' HEMS and DMS which possess multiple self-* capabilities. Thus, ODiS seeks answers to the following question: How can we create the most appropriate models, techniques, and algorithms to develop novel kinds of self-configuring, self-organising, self-healing, and self-optimising DMS that are integrally coupled with the distributed HEMS? In this article, the vision of ODiS is presented in detail based on a thorough review of the state of the art.
}

\dn\dn

%
\section{Introduction}

Climate change and sustainability are the major drivers to decarbonisation in the energy supply system by making use of renewable energy power plants (solar and wind energy, etc.). This goes hand-in-hand with increasing efficiency from primary energy to end-use energy and motivates shifting in the heating sector to heat pumps and in the traffic sector to electromobility. The power system is expected to become the backbone in all energy sectors and consequently, the fundamental critical infrastructure.
Components such as photovoltaic (PV) systems, heat pumps, electric vehicles, and stationary battery systems, are connected to the electrical distribution system. This results in a decentralisation of power supply and requires the coordination of millions of components owned by network customers who become more and more prosumers with HEMS.
From a distribution system operator's point of view, this transformation changes the operational approach of the distribution system completely. Presently, few sensors allow transparency about power flows and voltage profiles in the grid. The dominant design principle of the distribution system are expectations of maximum power injected or absorbed by customers. However, increasing reinforcement costs push system operators to operate the electrical grid more at its capacity limit. Monitoring and state estimation become necessary to gain transparency to know the real safety margin towards the capacity limit. Also, customers striving for self-sufficiency become more active in controlling their power flows to the public distribution grid, resulting in more volatile power flow patterns which increases requirements of monitoring and further control.

Operating the power system near to its capacity limit motivates to use flexibilities such as active and reactive power control capabilities of controllable \ac{DG}, loads, and storage systems. 
The coordination of these millions of actors in hundreds of thousands of grids can only be realised by full automation, but a large-scale rollout of automation technology in many countries is many years away due to various reasons:
\begin{shortitemize}
    \item Conventional regulatory schemes incentivise grid reinforcement instead of using \ac{ict}.
    \item Full transparency in low and medium voltage grids is non-existent at present. Over the next years near to full transparency is planned on the medium voltage level, on the low voltage level, this will be even farther in the future.
    \item The smart meter rollout is delayed due to information technology (IT) security requests and a limitation to large customers in the first rollout phases.
    \item A resilient and continuously available communication infrastructure to connect all sensors and actors is still to be built.
    \item Present ``smart energy'' products request significant effort to configure, parameterise, connect devices, etc.
    \item It is not yet entirely clear if from a full system perspective this shift is beneficiary about costs, reliability, resiliency, etc.
\end{shortitemize}

Solving the technical challenges paves the way for full automation of electrical distribution systems in interaction with \ac{hems} in the future. The goal is to construct an automation architecture that is as robust, safe, flexible, and trustworthy as possible. To achieve this goal, the electrical distribution operation architecture has to act more independently, flexibly, and autonomously. Such systems are also called ``organic'' after the term ``Organic Computing'' \cite{MST17,TomfordeSM17} because they should adapt dynamically to exogenous and endogenous change. The systems are characterised by the properties of self-organisation, self-configuration, self-optimisation, and self-healing based on self awareness as well as context awareness. In this sense they can also termed to be ``cognitive''. The ODiS (Organic Distribution Systems) initiative, therefore, seeks answers to the following question: 

\textbf{How can we develop the most appropriate models, techniques, and algorithms to create novel self-configuring, self-organising, self-healing, and self-optimising \ac{dms} that are integrally coupled with the distributed \ac{hems}?} 

Due to their ``organic'' nature, these systems will be called Organic DMS (O-DMS) and Organic HEMS (O-HEMS).

The remainder of this article is organised as follows: Section~\ref{sec:sota} summarises and assesses the current state of the art with a major focus on self-* capabilities in power grids. Afterwards, Section~\ref{sec:objectives} introduces the key research questions of the ODiS initiative, derives a system model allowing for self-* capabilities and discusses use cases as well as evaluation scenarios. Finally, Section~\ref{sec:conclusion} summarises the concepts and gives an outlook to the future work of the ODiS initiative.

\section{State of the Art}
\label{sec:sota} 

The smart grid architecture can be divided into layers, as Refaat~et~al.\ point out in \cite{2018_self-healing_control_strategy}: On the one hand, the \textbf{power system layer} is responsible for transporting electrical energy from generators to consumers. This layer is made up of electrical lines, transformers, power switches, and other assets used directly to transport electrical energy. 
On the other hand, the \textbf{communication layer} is responsible for the transmission of relevant data to operate the power grid. It is used to collect instrumentation and measurement data from various measuring devices and transfer them in a fast and reliable manner. The data can be transferred and used by a central instance, i.e., a grid control centre, or used by distributed data collectors in the smart grid paradigm. This layer may be a virtual private network on the public Internet or forms of local communication networks (e.g., WiFi).

The distribution management system (DMS) is a system that makes use of the communication layer to receive and transmit information to other entities (here modelled as \textit{agents}) in the grid. The goal of the DMS is to keep the power system layer in secure and cost-efficient operation conditions. The DMS interconnects with \ac{hems} situated in the corresponding power system area. 

\textbf{Advancing on the concepts of DMS and \ac{hems} as part of an \ac{ODiS}, both an O-DMS and an O-HEMS would possess several self-* capabilities which enable them to adapt and optimise for different situations.} The following subsection introduces the relevant terms to describe DMS and HEMS. Afterwards, the state of the art in self-*-capable distribution management systems is reviewed.

\subsection{Home Energy Management Systems and Distribution Management Systems}


Home Energy Management Systems (HEMS) aggregate information and manage the assets connected to it in a home/building. They can optimise the internal energy consumption and provide an economic advantage to the homeowners: 
Althaher~et~al.\ present their HEMS controller in \cite{2015_automated_demand_response}. The goal is to control appliances in response to dynamic price signals so that the electricity bill and the curtailed energy are minimised. Results show that both goals can be achieved given certain flexibility on the user's comfort. 
In \cite{2017_optimized_hems}, Ahmad~et~al.\ present a HEMS to minimise the electricity bill of the prosumer by optimising the energy consumption. The optimisation should also be beneficial for the utility due to the reduction of peak power. 
Zhang~et~al.\ present a demand response strategy for a HEMS that categorises loads in a household into three categories depending on their controllability \cite{2016_learning-based_demand_response}. The \ac{ML} based strategy uses data such as weather forecast and thermostat settings to optimise energy consumption.

Enhanced with self-*-capabilities, O-HEMS will also play a part in organic DMS. 
Little research has been done to use HEMS beneficially for the grid operator: Zandi~et~al. \cite{2018_implementation_hems} show the implementation of a HEMS as part of a \ac{MAS}. Three self-optimisation use cases are provided that could benefit from such a HEMS: First, the HEMS can be used to reduce peak load in the distribution grid as requested by the DMS. Second, it is used to reduce the voltage impact caused by high renewable feed-in. The third use case tries to improve residential resiliency by enabling islanding of the distribution grid while relying on local energy storage. 
Vázquez-Canteli and Nagy review numerous papers using reinforcement learning (RL) for demand response in HEMS \cite{2019_canteli_naga_rl_hems_review}. They identify multi-agent RL, dynamic addition and removal of buildings, and testing RL control systems in physical systems (possibly in the co-simulation context) as open research questions (among others). Similarly, Dimeas~et~al.\ propose a detailed concept of HEMS and envision services to assist grid operators in \cite{2014_smart_houses_kok}, but do not present practical research or simulation results.

A major innovation in research intended by  \ac{ODiS} is the use of several self-* capabilities that allow both an O-DMS and an O-HEMS to select control strategies and exploit flexibilities according to their individual and joint optimisation goals. This is believed to enhance flexibilities in changing situations and configurations and improve resilience to a variety of disturbance scenarios.

\label{ssec:odms}
A Distribution Management System (DMS) is defined by possessing different characteristics \cite{2009_epri_cigre_dms_workshop}: The system actively monitors the conditions in the distribution system; it can perform controlling measures in real-time (e.g., reconfiguration after faults; voltage and reactive power management). Additionally, it can integrate \ac{DG}, storage, and demand response. The assets can be integrated directly or aggregated in underlying HEMS. The HEMS can distribute aggregated measurements of its underlying assets and additionally it can send information about the current flexibilities to the DMS.

A DMS has to function in the different states a power grid can be in. According to \cite{abur2004power}, in a \textbf{normal grid state}, all loads are supplied with power and the grid is operated within the permitted conditions. In an \textbf{emergency state}, some operating limits are violated, which requires immediate action by the grid operator or the DMS to come back to a normal grid state. The \textbf{restorative state} appears if not all loads are supplied with power (e.g., partial or total blackout). Measures must be taken to resupply all customers while adhering to the defined operational limits \cite{2012_smart_grid_fundamentals}.

\textbf{Islanding} can appear in power systems and needs to be managed by the DMS. It can both be unintentional (as part of the restorative grid state), but also intentional\,/\,planned to operate parts of the grid autonomously. Islanding is defined as a \textit{``condition in which a portion of an Area} [sic] \textit{\ac{EPS} is energised solely by one or more Local \ac{EPS}s through the associated points of common coupling while that portion of the Area \ac{EPS} is electrically separated from the rest of the Area \ac{EPS}.''} \cite{2003_ieee_std_1547}. 
In other words, islanding is the formation of a power grid by \ac{DG} if the main grid is in a state of blackout. During intentional islanding, the distributed grid-forming generator performs voltage and frequency control to supply the connected loads with energy while respecting key performance indicators such as voltage and frequency bands. Conscious islanding of autonomous parts of a grid has been 
introduced as adaptive islanding or \textit{self-islanding}~\cite{yeager_smart_power_technologies_2005}.

The parts of the grid operated autonomously in this way are also referred to as \textbf{Microgrids} (MGs).
MGs are defined as distribution systems consisting of distributed loads and generation units operable ''in a controlled, coordinated way'' both in grid-connected and islanded mode \cite{marnay2015microgrids} and thus can be understood as a part of active distribution grids.
The authors of \cite{marnay2015microgrids} further categorize three types of MG controllers, namely centralized, distributed, and autonomous control, and point out the superior characteristics of distributed approaches regarding robustness and privacy concerns.

Many published works listed in the following section incorporate some form of self-* capability in their DMS design. However, none of the cited publications combines all of the introduced self-* capabilities into a ``fully'' organic DMS that can flexibly adjust to changing situations in a grid, both on the electrical and the IT side. 


\subsection{Methodical Foundations from the Fields of Autonomic and Organic Computing}

Orthogonal to the application area of the smart grid, self-* properties have been considered in the literature as an own research field. In the following, we define the desired self-* properties according to the terminology used in the Organic Computing (OC) \cite{MST17} and Autonomic Computing (AC) \cite{KC03} domains. However, there is no commonly agreed notation of these self-* terms in literature -- which is in turn not fundamentally necessary for the scope of the initiative as we aim at the basic functionality rather than a precise terminology framework. We assume a self-* system to consist of a potentially large set of (autonomous) subsystems. Internally, each of these autonomous subsystems distinguishes between a productive part (responsible for the basic purpose of the system) and a control mechanism (CM, responsible for controlling the behaviour of the productive system and deciding about relations to other subsystems). This corresponds to the separation of concerns between \textit{System under Observation and Control} (SuOC) and \textit{Observer/Controller} tandem in the terminology of OC \cite{TP+11} or \textit{Managed Resource} and \textit{Autonomic Manager} in terms of AC \cite{KC03}. The user guides the behaviour of the autonomous subsystems and the overall system depending on the abstraction level using a utility function and (normally) does not intervene at the decision level. Actual decisions are taken by the productive system and the CM.

According to \cite{KC03}, \textbf{we define \textit{self-configuration} as the ability of a CM to change the parameterisation of components and systems }following high-level policies. In contrast,\textbf{ \textit{self-organisation} is the ability of distributed controllers to ``modify the overall system's structure} (i.e., relations between components and the corresponding interaction schemes) depending on current conditions and based on the particular system goal'' \cite{MST17}. This can be augmented with \textbf{\textit{self-optimisation} capabilities,} which \textbf{means that the CM of the ``components and systems continually seek opportunities to improve their own performance and efficiency'' }\cite{KC03}. Considering robustness/resilience as a primary goal of self-* mechanisms, reactions of the CMs to disturbances or attacks are required. In this context, we distinguish between\textbf{ \textit{self-healing} (i.e., the CMs being able to ``automatically detect, diagnose, and repair localised software and hardware problems'' \cite{KC03})} and \textit{self-protection} (which refers to the ability of a system to ``automatically defend against malicious attacks or cascading failures. \textbf{}It uses early warning to anticipate and prevent system-wide failures.'' \cite{KC03}).

Mapping self-* terminology onto the power system field, the overall self-* system is represented by the organic distribution system while the autonomous subsystems comprise both, components of the O-DMS and the different O-HEMS. Productive systems in the O-DMS are the grid control strategies, which use flexibilities and actuators to keep the grid in secure conditions. The different CM and their realisation are subject to the work of the ODiS initiative. In the following, we discuss the four self-* capabilities that are investigated in \ac{ODiS}: a) self-configuration, b) self-organisation, c) self-optimisation and d) self-healing. For all four concepts, we introduce the term, describe the state of the art from an OC/AC perspective, and close with the state of the art in the domain of energy systems.

%

\subsubsection*{Self-Configuration}

{Self-configuration} (also called self-adaptation or re-parameterisation \cite{TomfordeG20}) in intelligent systems is either realised statically or adaptive. Static solutions are assumed to operate in fully predictable environments. In contrast, the main challenge is to develop solutions for adaptive self-configuration mechanisms, since this requires guarantees for reliability and accuracy despite the presence of uncertainty and unanticipated conditions \cite{KC03}. This includes consideration of a set of aspects \cite{OG+99}, including i) the type of adaptation (i.e., response to external behaviour, or closed systems), ii) the degree of autonomy, iii) cost-effectiveness, and iv) the selection of the currently 'best' adaptation strategy, see \cite{IC14-decisionmaking}. Furthermore, categorisation with respect to real-time capabilities for embedded control mechanisms is used in the literature \cite{BP2009examination}.

In general, there are three categories of decision-making techniques \cite{IC14-decisionmaking,KRUPITZER2015184}: predefined decisions, decision making driven by ML, and constraint programming. Predefined decisions mean that the behaviour for a specific situation is specified at design-time by engineers, ML approaches to integrate the self-configuration decision within the learning algorithm or use learning techniques to derive predefined decisions, and constraint programming tries to find an optimal solution using solvers. Due to the dynamic nature of the control problem focused in \ac{ODiS}, we will rely on ML approaches.

Specific to power systems, there is only limited research on self-configuring systems. Currently, the project ``Plug'n'Control'' 
aims to validate the concept of a self-configuring system of inverters. By allowing inverters to detect their underlying asset (battery, charging station etc.) and being able to communicate, the system can be used to avoid overloading in the grid or perform other tasks to retain grid stability. It is unclear which category of decision-making is used.

To our knowledge, no (organic) distribution management system uses ML-based decision-making. Therefore, the initiative provides an ideal testbed for adapting and evaluating recent advances in ML in terms of their use in self-configuring distribution grid management systems.


\subsubsection*{Self-Organisation}

{Self-organisation} has been defined as ``a  mechanism  or  a process which enables a system to change its structure without explicit command during its execution time'' \cite{serugendo2005self}. Based on a multi-agent perspective and according to \cite{ye2016survey}, the different concepts can be classified using the main categories of task or resource allocation, relation adaptation, organisational design, and collective decision making (next to some minor classes). 

Self-organising task allocation has been primarily studied as a basis for other research issues such as coalition formation \cite{shehory1998methods}. In addition, it heavily relies on the particular application scenario and the corresponding purpose of the allocation -- the most prominent algorithms have been considered in the context of wireless sensor networks \cite{low2004task} and multi-robot systems \cite{liu2007towards}. In all of these cases, specific tasks of the sensor nodes or the robots have been assigned to the individual entities using decentralised algorithms. Although being based on standard communication schemes for distributed systems (such as leader election, echo algorithms, or broadcast \cite{tanenbaum2017distributed}), the solutions are not generic.

Relation adaptation typically covers two intertwined questions for the autonomous subsystems \cite{TomfordeKS17}: with whom to modify relations and how to modify them. Existing approaches for partner selection are based on the agents' interaction history \cite{yan2013survey}, sometimes combined with an uncertainty or trust measure \cite{EdenhoferTKKBHM15}. In contrast, the term 'organisational design' refers to the question of how society or group members interact and relate with one another \cite{horling2008using}. This mainly means work focusing on the assignment of roles to different actors or agents in the overall system \cite{galbraith2012evolution}. In contrast, we assume given roles in the context of \ac{ODiS}.

The concept of collective decision making originates from social science, where members of a group have to achieve a consensus that typically reflects a compromise between the different members. As one particular example, Valentini et al.\ \cite{valentini2014self} presented a generalised solution that assigns weights to members and considers their votes in a decentralised manner. In general, the idea is mostly to find a collective agreement over the most favourable choice among a set of alternatives. This is usually encoded as the 'best-of-$n$' decision problem, which is characterised by a dynamic $n$ in open environments. For decentralised decisions, we will investigate how these decision mechanisms can be reused for energy systems. 

Self-organisation in power systems is only a fringe topic in research, mostly implicitly researched in the form of cooperative \ac{MAS}. 
A comprehensive survey on MAS in power system control can be found in \cite{kantamneni2015survey, tazi2019multi}.
For example, Jianfang~et~al. develop a \ac{MAS} for self-healing in \cite{2016_hierarchical_control_model}, which relies on self-organisation for fault isolation and resupply.  
In \cite{niesse2012market}, an agent-based control scheme for the trading of active power and ancillary services is proposed that guarantees the fulfillment of grid constraints.
Dimeas et al. apply MAS to MGs, where each of the following elements is represented by an agent: agent network operator, market operator and central controller of the individual MGs \cite{dimeas2005operation}. The considered actions comprise load-shedding, black start and the transition between grid-connected and islanded mode.
In the context of {self-healing}, Li et al. develop an agent-based fully decentralized framework for service restoration of a distribution grid with peer-to-peer communication between the agents \cite{li2019full}. In the presented strategy, intentional islanding is combined with network reconfiguration. 
However, the vision of ODiS goes far beyond these approaches by including self-organisation only as one of many self-* capabilities.



\subsubsection*{Self-Optimisation}

{Self-optimisation} in self-adaptive and self-organising (SASO) systems is typically considered to deal with the improvement of control parameters of the productive system or the management of the overall system structure based on a given set of goals. Depending on the time horizon, two different fields are distinguished:  ML-based and optimisation-based.

In~\cite{SASO_Planning-as-Optimization}, the usage of optimisation techniques within SASO systems for generating new system configurations or adaptation plans is analysed. The authors identified the use of 29 different techniques in 51 publications, including techniques from probabilistic, combinatorial, evolutionary, stochastic, mathematical, and meta-heuristic optimisation. 
The approaches target optimisation in centralised systems with only minor attention to decentralised optimisation. Such centralised optimisation comes with the disadvantages of providing the risk of a single point of failure as well as a potential bottleneck. Further, it contradicts the nature of autonomous, coordinating entities. Besides applying optimisation heuristics based on a given utility function, constraint solvers are frequently applied. 

On the other hand, self-optimisation is achieved by autonomous learning. In a study~\cite{DAngeloGGGNPT19} of our preliminary work, we analysed the usage of learning techniques in SASO systems. Here, RL has been identified as a most prominent variant, mainly realised as simple learning tasks, e.g., using Q-Learning, or more sophisticated approaches, e.g., using Learning Classifier Systems~\cite{rizk2018decision}. Alternatively, multi-agent RL (MARL) is employed to solve problems in a distributed manner when centralised control becomes infeasible~\cite{Marinescu2017}. 
MARL is also used in~\cite{Malialis2016} to tackle the complexity emerging in \ac{MAS} domains. 
Especially the ``Extended Classifier System'' (XCS) variant by Wilson \cite{Wil95} (including its variants from the OC domain such as \cite{prothmann2009organic,fredivianus2012stay,SERH16,SRTH17}) has been widely used for implementing self-adaptation with runtime learning capabilities. For instance, XCS can be seen as an integral part of OC systems that are said to exhibit self-learning properties. 

Self-learning has not been a major focus in research on self-optimising power systems. While many publications limit their control capabilities to adjusting switching states, other control measures can be used to optimise the power system: 
Wong, Lim, and Morris publish a so-called ``self-intelligent'' active distribution management system in \cite{2013_self-intelligent_active_management_system}. Their system can control key grid variables such as voltage limits by using an energy storage system to mitigate issues in the grid.
The reliable and economic operation of a distribution grid is targeted by the optimisation structure presented in \cite{2016_economic_optimization_of_self-healing_control} by Li~et~al. Different asset classes (loads, PV generators, wind energy converters, batteries) are modelled as individual agents of a \ac{MAS} that can perform certain tasks not only for self-healing, but also for economic optimisation of the grid.
Roytelman and Medina present a volt/VAR control algorithm that uses advanced metering infrastructure measurements, distributed capacitors and inverters to stabilise the grid's voltage profile \cite{2016_volt_var_control_dms}.

Recently, there have been a few works focussing on cellular network structure approaches. 
E.g., the project Zellnetz2050 \cite{flatter2021zellnetz} follows an approach to increase the flexibility of future energy systems in which different sectors of the energy systems (electricity, gas, mobility and heat) are coupled. The cells are created hierarchically based on geographical regions and comprise all levels of the electricity grid. In contrast, the ODiS initiative focuses on the distribution level of the electricity sector, but also considers the ICT system and other grid states. 
Another cellular approach is developed in the project C/sells (\url{www.csells.net}). Based on autonomous energy cells that can be created both region- and asset-based and manage their own consumption and generation locally, a decentralized power supply is pursued. At the same time, a variety of methods for the operational management of the individual cells is enabled. In this context, the ODiS initiative also aims at answering the question of the communication structure in the automated distribution grid. However, similarly as in Zellnetz2050, C/sells does not provide strategies for events of emergency.

An approach from the Organic Computing domain has been presented by Schmeck et al.\ in the context of their 'Smart Home': A building energy management system (i.e., a HEMS) that combines different communication technologies for monitoring, data recording, and visualisation purposes \cite{BeckerKLMS15}. This information is used to perform offline optimisations using heuristics to improve the utilisation of flexibilities for goals such as an increased self-consumption or the provisioning of grid-supporting services. The system is based on a hardware demonstrator ('living lab') and is evaluated by hardware-in-the-loop simulation \cite{kochanneck2018hardware}. In contrast to \ac{ODiS}, the focus is mostly on isolated HEMS. In particular, the awareness level is less sophisticated, the configuration of the HEMS is more static, the relation to other HEMS is neglected, and the integration into the DMS is done rather static.

An agent-based controller based on RL achieves power balance in a complex smart grid with changing communication topologies \cite{singh2017distributed}. 
In their study, Rahman et al.\ propose a \ac{MAS} in which distributed agents estimate the reactive power based on local and information of their neighbours \cite{rahman2016agent}. Based on this estimation, controller agents adapt to the current system state and ensure the safe operation of the distribution network.
Radhakrishnan et al.\ present an agent-based energy management system in which the responsibilities for switching between operating modes, the commitment of storage units and operation schedules are assigned to different agent types \cite{radhakrishnan2016multi}.
In \cite{merdan2011multi}, a hierarchical multi-agent architecture performs the energy management of an active distribution grid while being flexible to changes in the grid topology. 
In further work, the authors also consider self-reconfiguration \cite{lepuschitz2010toward}, robustness \cite{vallee2011decentralized}, and self-diagnostics \cite{merdan2011monitoring}.


Current research usually focuses on one aspect of the grid to be optimised such as the switching states or energy storage. \ac{ODiS} aims to use a holistic approach to self-optimisation, assessing and weighting not only the different flexibilities and topologies but also the optimal control strategy in the specific scenario. Two different levels -- grid-level via O-DMS and home-level via O-HEMS -- provide a large action space to operate within -- greater than in most similar publications. A major challenge to research in \ac{ODiS} will be the reduction of the action space depending on the current situation in the grid.

\subsubsection*{Self-Healing}

The basic idea of approaches for {self-healing} following the initial notion of IBM's Autonomic Computing initiative aimed at decreasing the required degree of human interaction \cite{KC03,chess2004work}. Self-healing is often examined in combination with self-protection. However, self-protection is usually considered in a security sense, which is not initially
a focus of research in ODiS. According to \cite{schneider2015survey}, self-healing systems are categorised as being fully supervised, semi-supervised, or unsupervised. 
This refers mainly to the time and the trigger of performing self-healing mechanisms with a focus on their learning strategies: manually or automatically. Fully supervised techniques (such as \cite{garlan2004rainbow,ahmed2009self,simmonds2010monitoring}) typically require continuous availability of human operators and frequent interaction. This is expected to result in reduced uncertainty about system behaviour and controlled re-parametrisation towards stable behaviour \cite{de2004conflict}. Popular approaches utilise a database of predefined recovery plans, and these plans are mapped towards observed or estimated events. Examples are the ``rainbow'' framework \cite{garlan2004rainbow} and the ``GPAC'' (General Purpose Autonomic Computing) framework \cite{calinescu2009general}. However, this can be mapped on the concept of predefined self-configuration that is triggered by a given set of events that are associated with negative states requiring repair mechanisms.

As an alternative to supervised approaches, semi-supervised concepts have been proposed that work based on event-driven monitoring. In general, this means combining classification of conditions (i.e. detecting events) with a run time analysis to identify anomalies, i.e. distinguishing between normal or known/established patterns of behaviour and unknown or noticeable patterns \cite{GruhlSWTH15,GruhlST21}. In literature, different approaches are available, ranging from purely reactive usage of basic exponential smoothing algorithms for time series \cite{metzger2013accurate} to the pro-active prediction of states \cite{engel2011towards}. One particular example is the VieCure framework \cite{psaier2010behavior} that combines event detection with direct analysis of metrics. In contrast to approaches that directly map faults to recovery plans, this system tries to determine abnormal behaviour and uses this as a trigger for self-healing decisions. Abnormal behaviour can be, for instance, a series of incidents within a log file -- and this is correlated with faults and correspondingly, the most suitable response is determined. However, unknown events and faults require supervision and this is not performed as a fully autonomous process.

There are only a few cases where frameworks are autonomously establishing self-healing properties, see e.g.\ \cite{miorandi2009embryonic,ramirez2011plato,dean2012ubl}. Their realisation is usually highly dependent on the particular application scenario and only seldom follows a generalised approach. Technically, they make use of genetic algorithms \cite{ramirez2011plato}, neural networks \cite{dean2012ubl} or so-called ``totipotent'' behaviours \cite{miorandi2009embryonic}. Besides the underlying applied techniques, the approaches differ in terms of how much risk and resource commitment are needed, which again controls how much autonomous behaviour is possible.

Focusing on power systems, numerous self-healing schemes have been published in recent times. For example, Xiu~et~al.\ present their research on self-healing in smart distribution grids in \cite{2016_research_on_self_healing_technology}. They define several core processes of a self-healing system. They also define different layers for agents in a \ac{MAS}, similar to \cite{2016_hierarchical_control_model}. The \ac{MAS} can reconfigure the grid after a fault by changing switches to disconnect and connect lines.
In \cite{2016_substation-based_self-healing_system}, Duarte et al.\ present their self-healing system for power systems. Their modelled scheme includes state estimation and prognosis of the power grid based on SCADA \cite{2012_crastan_elektrische_energieversorgung} data, a self-healing optimisation and a sequence of manoeuvres to change switching states according to the optimisation result. In simulations, they show that the scheme can reduce the SAIDI \cite{2012_crastan_elektrische_energieversorgung} of different feeders, sometimes by 20\,\% or more. 
Similar fault detection, isolation, and restoration (FDIR) schemes are presented in \cite{2016_implementing_self-healing_distribution_systems} by Guo~et~al., \cite{2018_dynamic_reconfiguration_and_fault_isolation} by Arif, Ma and Wang in \cite{2016_a_novel_fault_self-recovery_strategy}, or \cite{2016_a_novel_fault_self-recovery_strategy} by Niu~et~al.
Liu~et~al.\ present a concept for a self-healing urban power grid in \cite{2012_the_control_and_analysis_of_self-healing_urban_power_grid}. It is based on a \ac{MAS}, which is structured in three layers (organisation, coordination, and response). The system classifies the state into 5 classes (secure normal, insecure normal, alert, emergency, restorative) and also defines 4 sub-control tasks: emergency control, restorative control, corrective control, and preventive control. Several test cases confirm the validity of the concept.
\cite{2007_software_agents_in_distribution_networks} by Baxevanos and Labridis details a cooperative system for fault and power restoration management in distribution grids. 

A common theme among published research on self-healing in power systems is the use of \ac{MAS}. Agents can coordinate with each other to perform self-healing. This is usually done after the occurrence of fault situations, where a new switching state of the grid has to be found \cite{2016_hierarchical_control_model}. A limitation of such schemes is the outage of parts of the grid if no connection to the substation transformer can be made by switching after a fault.

Self-healing can benefit from the coordination between O-DMS and O-HEMS levels. By making use of local control strategies and \ac{DG} as well as batteries, grid sections can be operated in a ``self-islanding'' mode without connection to the substation transformer while the fault is being repaired. Research on such a scheme is a major innovation compared to the state of the art.

%
%




%
%
\section{A Concept for Organic Distribution Management Systems}

\label{sec:objectives}

\subsection{Key Research Question}

The overall goal of ODiS is to develop and investigate a novel, integrated approach for an organic distribution system that operates completely autonomously. In particular, we have to answer the following basic research question: 

\textbf{How can we create the most appropriate models, techniques, and algorithms to develop novel kinds of self-configuring, self-organising, self-healing, and self-optimising DMS that are integrally coupled with the distributed HEMS?}

These new O-DMS and O-HEMS (``O'' for ``Organic'' to refer to ``Organic Computing'' as reference domain) shall be implemented as the management system for a benchmark grid and evaluated by simulations according to a number of use cases. The use cases are chosen in a way that the key self-* capabilities are required.

\begin{figure}[b]
    \centering
    \includegraphics[width=0.86\textwidth]{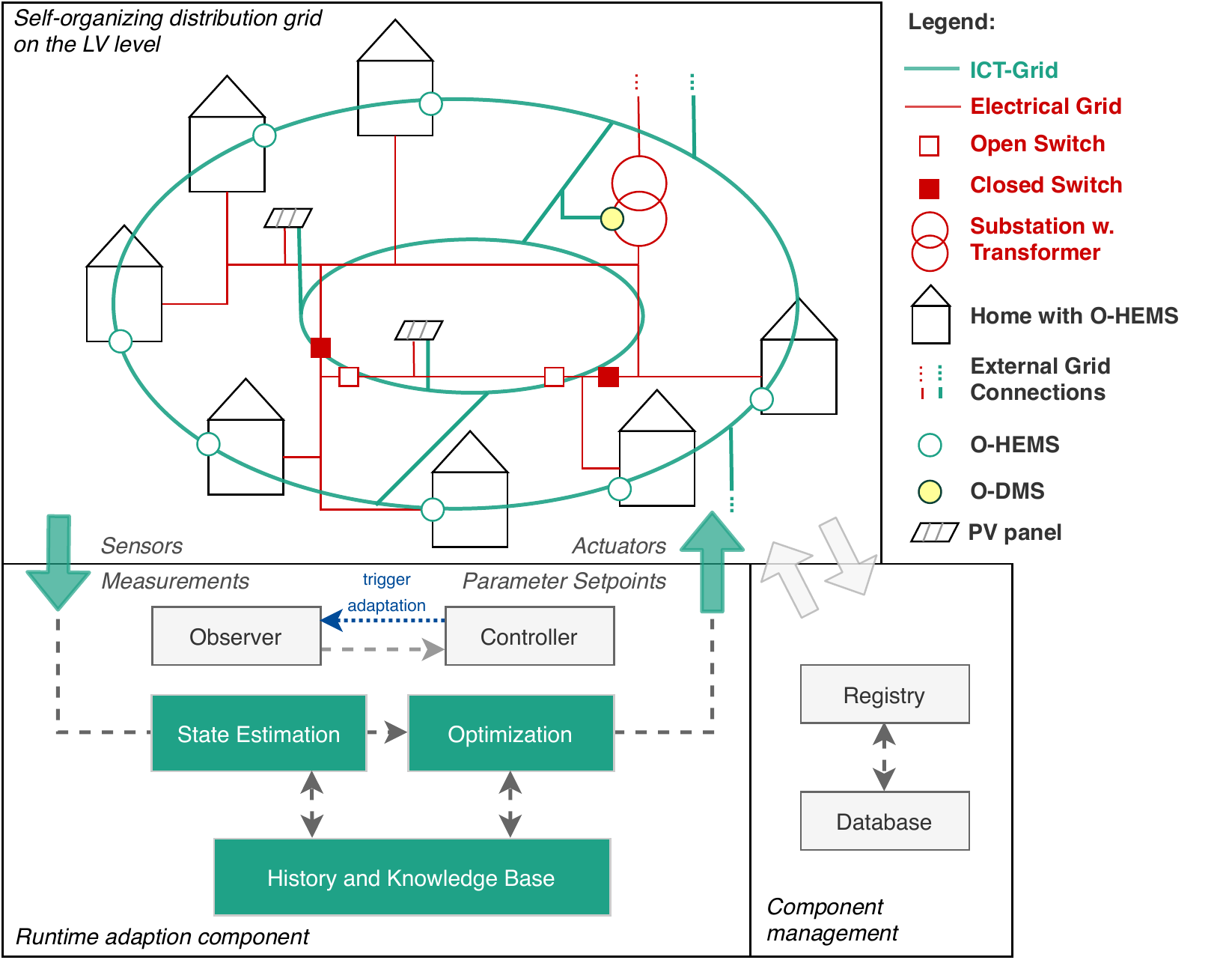}
    \caption{Self-* capable low voltage grid including the O-DMS and several homes with active components (O-HEMS, inverters, loads. ...). The red mesh depicts the electrical grid, which connects the various homes to the substation. Some open and closed circuit breakers (switches) create a radial topology. In the context of \ac{ODiS}, we assume that switches inside the cable distribution cabinet are remotely controllable. The green lines depict the communication network to which the O-DMS, various O-HEMS and circuit breakers are connected.} 
    \label{selfx_grid}
\end{figure}
 
The upper part of Figure~\ref{selfx_grid} shows a low voltage grid that is outfitted with an O-DMS. The lower part of the figure shows the runtime adaption components of the O-DMS. Sensors provide various measurements, which are the basis for the state estimation process (observer). The optimisation (controller) creates setpoints depending on the current state and the control strategy. 
Setpoints are sent to the actuators in the grid.
    
\begin{figure}
    \centering
    \includegraphics[width=0.84\textwidth]{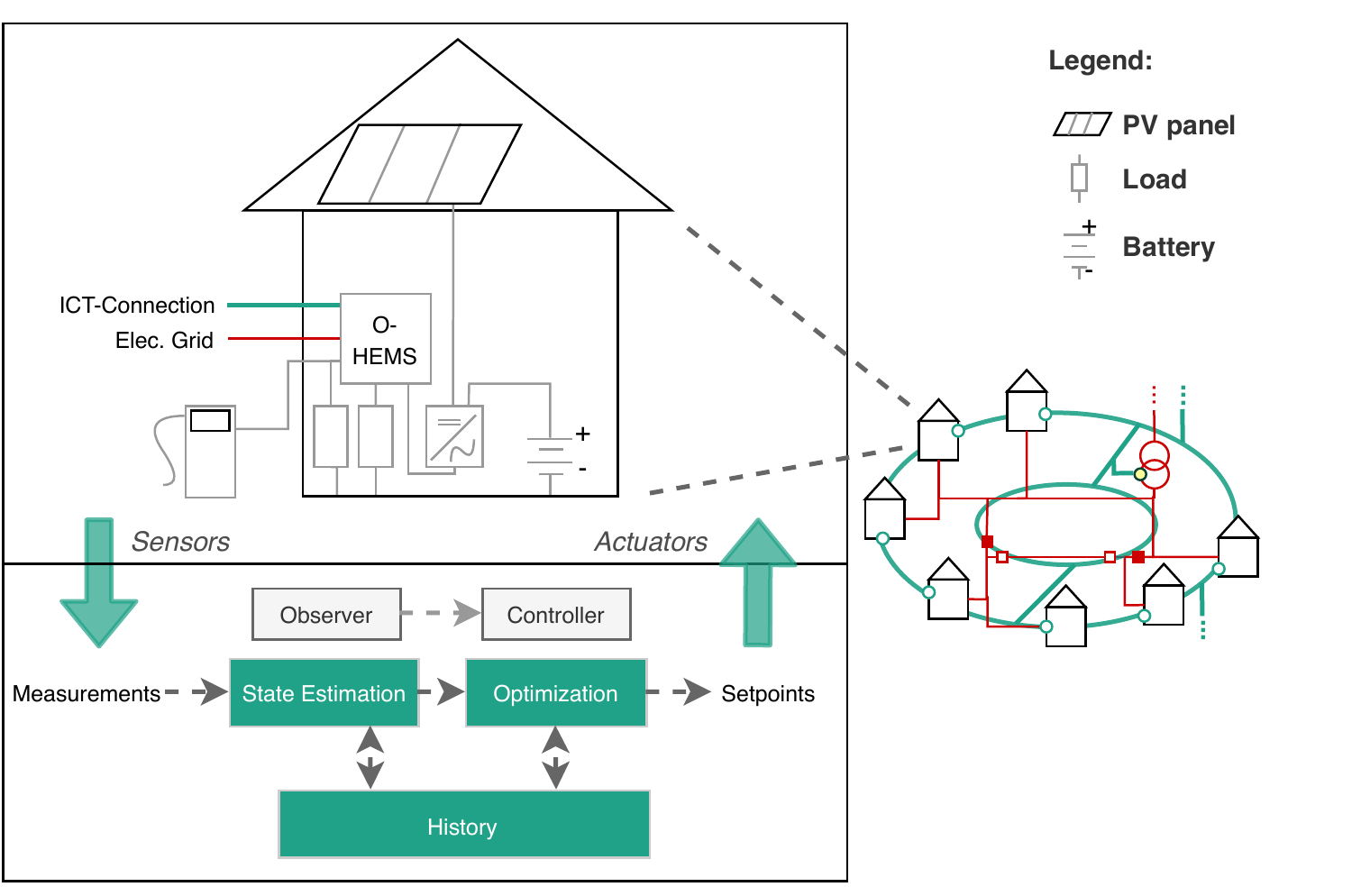}
    \caption{Zoom into a single home which -- among the O-HEMS -- includes components for the generation, consumption, and storage of energy (e.g., household loads (resistances), PV panels and battery with an inverter, electric car charger etc.).} 
    \label{selfx_home}
\end{figure}

Figure~\ref{selfx_home} zooms into Figure~\ref{selfx_grid} to reveal the components of an exemplified home. At the center, the O-HEMS connects to the ICT and the electrical grids. The O-HEMS can act as an agent with underlying assets such as inverters are part of local control strategies. The runtime adaption component is identical to the O-DMS. However, state estimation and optimisation blocks use different functionality due to the different scope of operation and available measurements and actuators. 

\subsection{System Model} 
\label{ssec:opsim_system_model}

\textit{Simulation framework}

\begin{figure}[b]
    \centering
    \includegraphics[width=0.8\textwidth]{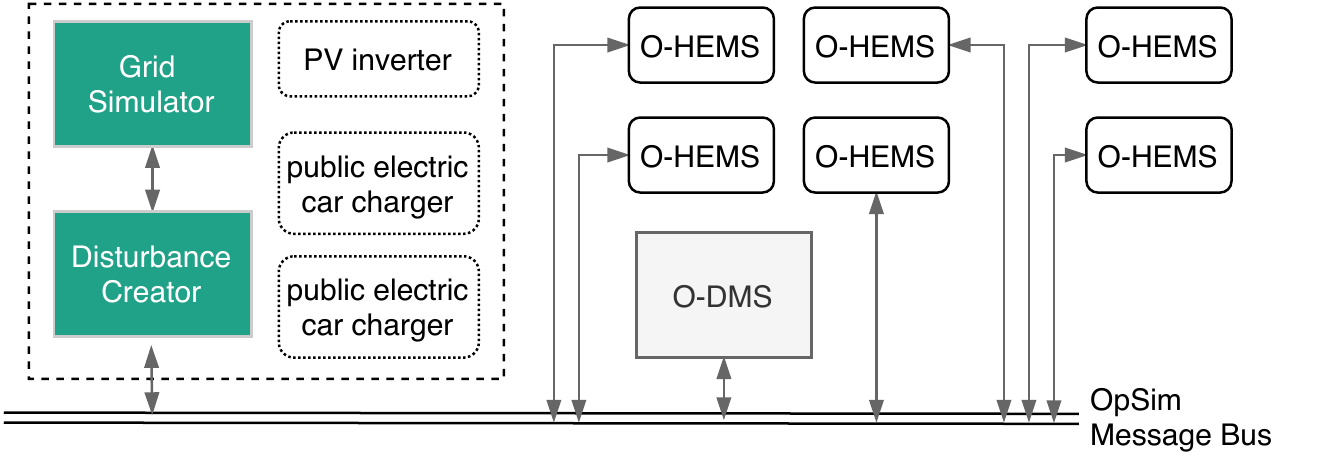}
    \caption{System model of the simulation framework operated as an \ac{MAS}.}
    \label{system_model}
\end{figure}

Figures~\ref{selfx_grid} and \ref{selfx_home} show the descriptive system model for the two relevant layers (grid level and home level). The system model is shown in Figure~\ref{system_model} and is visualised within the simulation framework \textit{OpSim} \cite{2018_opsim_abschlussbericht} which enables the exchange of information in a predefined way. All actors as well as a grid and communication network simulator connect to a simulation bus, which routes the information automatically.
\textit{OpSim} has already been used in a variety of applications, e.g., for co-simulation between different types of components, testing of prototype controllers and other tools used to aid grid operation. It consists of a ``master control program'' and a message bus to which components can connect. Using a scenario configuration, the control software routes information between the different components. 

For ODiS, \textit{OpSim} can be utilised to simulate various scenarios in different use cases.
All agents (the term is used for either O-DMS and O-HEMS) connect to the message bus as individual \textit{OpSim} components. Depending on the available functionality, the runtime adaption component is run in parallel for every component. Measurements are observed, processed and setpoints are generated. The setpoints are sent back to the message bus and distributed.
A separate component holds the grid model and the corresponding time series. It is used to generate measurements for the other components. The disturbances imposed by the use case are created in the ``disturbance creator'' component by altering the data going out to the message bus accordingly. Part of the research is to identify and evaluate ways to realistically emulate disturbances for the corresponding use cases. The grid simulator also collects the set points by the O-DMS, O-HEMS so that the next grid state is a reaction to the chosen control strategies.

\vspace{1ex}
\textit{Observable and controllable parameters}

\begin{figure}[b]
    \centering
    \input{figures/InputOutput_ODMS_labels_new.tex}
    \caption{Schematic system model of the \ac{ODiS}. 
    The trajectory $\mathbf{X}^k$ of a variable $X \in \mathbb{R}$ 
    contains all values $\mathbf{X}^k=[X(k|k) X(k+1|k) \ldots X(k+H|k)]$ computed at the current time step $k \in \mathbb{N}$ over a time horizon $H \in \mathbb{N}$ where $X(k+h|k), h \in \{1,\ldots,H\}$ corresponds to the value of $X$ at time step $k+h$ computed at time step $k$. For simplicity, we denote $X(k|k)$ by $X^k$. $\overline{X}$ is the maximum value of $X$, $\underline{X}$ the minimum value of $X$ and ${X}^\star$ the setpoint value of $X$. 
    }
    \label{inputoutput}
\end{figure}
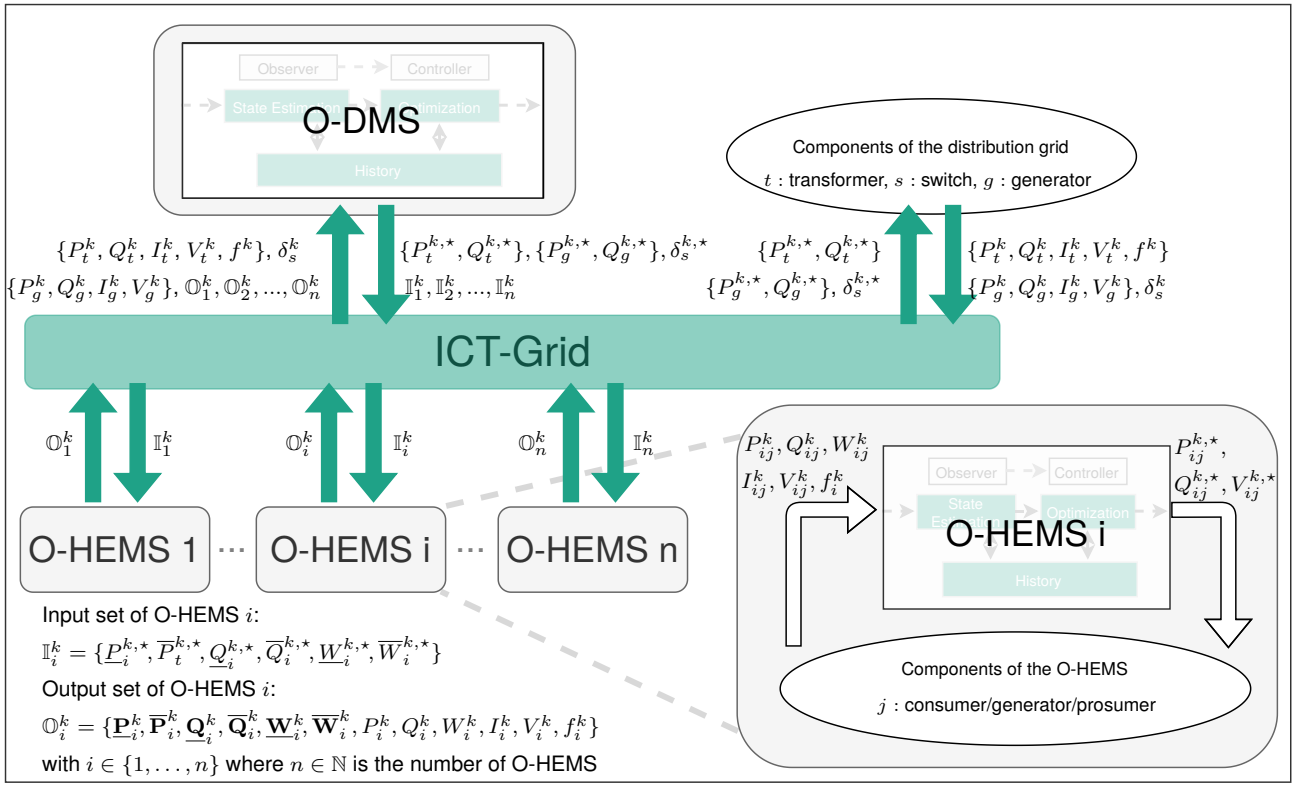
Every component observes different variables and adjusts parameters depending on the self-optimisation targets as shown in Fig.\ \ref{inputoutput}: 
The \textbf{O-DMS} component has access to measurements from the corresponding power grid and the data from underlying components of the distribution grid. 
Specifically, the O-DMS reads measurements of active power ($P_{t}^k$, $P_{g}^k$), reactive power ($Q_{t}^k$, $Q_{g}^k$), current ($I_{t}^k$, $I_{g}^k$), voltage magnitude ($V_{t}^k$, $V_{g}^k$), and  frequency ($f^k$) from the substation transformer $t$ and from \ac{DG} $g$ connected at the O-DMS voltage level respectively. 
Further, circuit breaker states (switching states $\delta_{s}^k$) are stored. 
From each O-HEMS the O-DMS receives both measurements and trajectories of flexibilities of relevant variables, such as the active and reactive power, storage energy capacity, current, voltage magnitude and frequency. 
From each O-HEMS $i$, the O-DMS receives measurements of the aggregated active power $P_i^k$, reactive power $Q_i^k$, the total storage energy capacity $W_i^k$, the current $I_i^k$, voltage magnitude $V_i^k$ and the frequency $f_i^k$ at the grid connection point of the O-HEMS.
Additionally, each O-HEMS $i$ transmits trajectories of flexibilities of active power ($\underline{\mathbf{P}}_i^k, \overline{\mathbf{P}}_i^k$), reactive power ($\underline{\mathbf{Q}}_i^k, \overline{\mathbf{Q}}_i^k$), and storage energy capacity ($\underline{\mathbf{W}}_i^k, \overline{\mathbf{W}}_i^k$), where $k \in \mathbb{N}$ is the current time step and $\mathbf{X}^k=[X(k|k) X(k+1|k) \ldots X(k+H|k)]$ the trajectory of a variable $X \in \mathbb{R}$ 
computed at $k$ over a time horizon $H \in \mathbb{N}$. 

This input data is then processed by the O-DMS to estimate the current state and to generate setpoints to control the grid state. 
The O-DMS can control transformer tap positions for voltage regulation by setpoints for active and reactive power $P_t^{k,\star}$ and $Q_t^{k,\star}$, respectively. Similarly, setpoints $P_g^{k,\star}$ and $Q_g^{k,\star}$ are computed for generators (e.g., solar panels) directly connected to the distribution grid. It also computes setpoints for the circuit breaker status $\delta^\star_s$ of each circuit breaker $s \in 
\mathbb{N}$ in the corresponding grid section to change the grid topology. Finally, it sends parameter ranges for active power $\underline{{P}}_i^{k}\!, \overline{{P}}_i^{k}\!$, reactive power $ \underline{{Q}}_i^{k}, \overline{{Q}}_i^{k}\!$ and storage energy capacity $\underline{{W}}_i^{k}\!, \overline{{W}}_i^{k}$ to the underlying O-HEMS. 

The \textbf{O-HEMS} 
gathers several measurements, depending on the connected assets: 
Each connected prosumer (e.g., electrical batteries) transmits its available energy. 
Flexibilities 
regarding loads are also calculated. 
The measurements are processed locally for use in local control strategies. 
If an O-HEMS is acting locally or provided with parameter ranges, it can self-optimise and set parameters to the inverters (for reactive power management), battery (for power storage/generation) or loads depending on available flexibility. 
%
Based on the measurements and forecasts, the O-HEMS aggregates its flexibilities of total storage energy capacity, active and reactive power for the current time step and for a predefined time horizon. The corresponding trajectories of active power, 
reactive power 
and storage energy capacity 
are transmitted to the O-DMS. 


For simulations in ODiS, all measurements can be assumed to be sent periodically in varying intervals which range from seconds to minutes depending on the component. For more specific use cases such as self-healing, the parameters need to be extended.


\subsection{Evaluation} 

To quantify the success of ODiS before actually implementing the techniques in real-world applications, various use cases as proposed in this section must be investigated.

%
\label{ssec:use-cases}

\begin{figure}[b]
    \centering
    \includegraphics[width=0.75\textwidth]{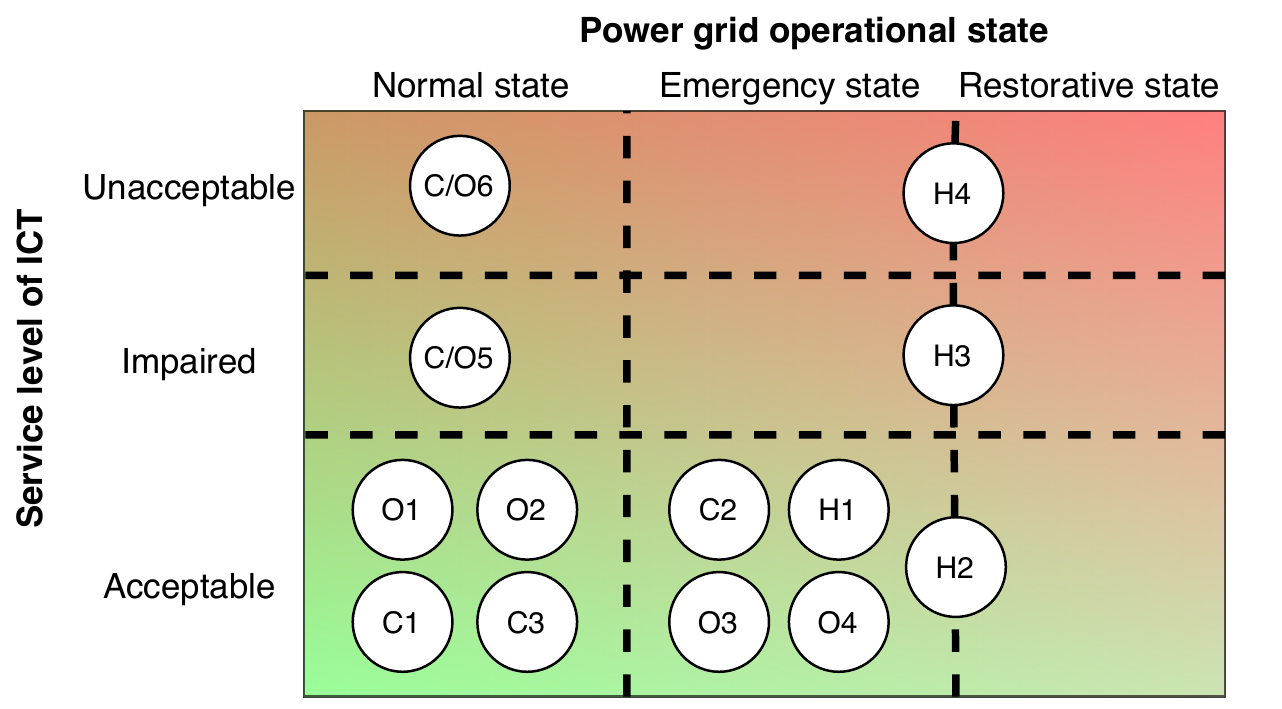}
    \caption{Self-* capabilities with use cases mapped to a two-dimensional state space based on \cite{sterbenz2014redundancy}. A high ICT service level is defined by high availability, accuracy, and low latency in the network. An impaired ICT service level entails communication issues (e.g., high latency, packet loss), making the data stream less reliable, while unacceptable quality is caused by the outage of some communication network resulting in complete loss of data (for a certain time). The power grid operational state is introduced in Section~\ref{ssec:odms}.} 
    \label{grid_state_to_use_case}
\end{figure}

The research in ODiS is categorised along with several self-* capabilities, that should be part of both the O-DMS and the underlying O-HEMS. The self-* capabilities are designed to ensure normal operation under a variety of interference influences both on the side of the electrical grid and of the communication network. The following summary of use cases is classified into mostly self-configuration/-organisation (C1-C3), self-optimisation (O1-O4, C/O5, C/O6) and self-healing (H1-H4). All use cases are described in Table~\ref{tab:use_cases} and sorted according to service level and power grid operational state in Figure~\ref{grid_state_to_use_case}.

\begin{table}
\centering
\small
\renewcommand{\arraystretch}{1.15}
\caption{List of use cases for the Organic Distribution System. Use cases containing \textbf{C} mainly require the ability to self-configure, \textbf{O} the ability to self-optimise, and \textbf{H} the ability to self-heal. Self-organisation is mainly included in the use cases containing \textbf{C}.}
\begin{tabularx}{\linewidth}{|@{\hspace{.2em}}>{\bfseries}l@{\hspace{.5em}}|X|}
\hline
C1 & \textit{Manual (Dis)connection of a new electrical asset:} Self-configuration is required after detection of new assets, as well as self-configuration of the grid model, operational algorithms, local parameters, communication configuration. The new power grid topology needs to be validated based on self-organisation. \\ \hline
C2 & \textit{Continuous anomaly detection, automatic background process:} The system needs to continuously monitor the data streams for anomalies and inconsistencies as could be caused by mal-parameterisation, human failure, or cyber-attacks. If anomalies are detected, the underlying data or control schemes need to be adjusted accordingly. \\ \hline
C3 & \textit{(Dis)connection of a new agent:} Every O-HEMS or O-DMS (agent) must perform self-configuration to adapt to the new overall system. This further requires self-organisation for determining the best possible system structure. \\ \hline
O1 & \textit{Self- and environment-awareness (including state estimation and forecasting):} The basis for optimisation of electrical grids is the knowledge of the relevant grid variables. A state estimation ensures that such knowledge is available, even when a low number of real-time measurements is available, which is typically the case for distribution grids. Forecasting provides projections into the future, which can be strongly beneficial for control strategies. In principle, both, O-HEMS and O-DMS, must be self- and environment-aware. \\ \hline
O2 & \textit{Normal operation and self-optimisation:} During normal operation, key performance indicators are inside their defined boundaries. Nevertheless, self-optimisation of agents can improve their conditions further, e.g., to minimise system losses or increase self-consumption of energy. \\ \hline
O3 & \textit{Voltage limit violation (no circuit breaker tripping):} High load, high \ac{DG}, or a sub-optimal switching configuration can cause voltage limit violations in a grid zone. Self-optimising agents need to choose between different measures to push the voltage back inside limits. Measures can be self-optimisation via load management, active and reactive power control of \ac{DG}, flexibilities/storage, switching state reconfiguration, or transformer tap adjustments. \\ \hline
O4 & \textit{Thermal line/transformer limit violation (no circuit breaker tripping):} The current flowing through grid assets is bound by thermal limits, which should not be violated for longer than a specific period or failures can occur. Similar to the use case O2, agents need to be aware of potential limit violations and use available measures to alleviate the limit violation. \\ \hline
C/O5 & \textit{Unreliable communication for the asset(s):} Communication between the agents can become unreliable, e.g., faulty measurement devices produce unreliable measurements that should not be used further. The communication network can become unstable, e.g., exhibit packet loss. Agents need to identify unreliable data streams and self-adapt accordingly. \\ \hline
C/O6 & \textit{Interrupted communication for the asset(s):} If technical issues appear in the communication network used for self-coordination, the exchange of information becomes unreliable or interrupted completely. Similarly to C/O5, agents need to self-adapt, e.g., by using an alternative channel of communication or by changing their control scheme from central to local control strategies. \\ \hline
H1 & \textit{Self-protection:} In the case of an asset failure or a manual misconfiguration, events such as blackouts can occur. For such cases, a self-protecting O-DMS prepares contingency strategies so that a minimum number of customers are affected.\\ \hline
H2 & \textit{Self-healing/reconfiguration:} If the self-protection is not successfully performed, a self-healing and reconfiguration scheme must be run. The scheme can trigger individual parts of a grid to operate as autonomous islands without connection to the higher voltage level (self-islanding) and may require changes in control strategies of O-HEMS etc. (e.g., switching to local control strategies). The process includes decoupling from the interconnected grid, a black start, islanding operation with suitably control strategies and re-synchronisation when self-islanding is no longer optimal.
\\ \hline
H3 & \textit{Self-healing under impaired ICT:} 
Impaired ICT connections pose an additional challenge for self-healing schemes. O-DMS and O-HEMS need to evaluate the limited control options and prioritise operational limit violations, e.g. allow for thermal overloading up to a certain time frame to avoid a (partial) blackout.
\\ \hline
H4 & \textit{Blackout in the communication grid:} A blackout in the communication grid prevents coordination among the different agents. The O-DMS needs to evaluate and activate alternative means of communication. If that is not possible, the agents need to switch to local control strategies automatically. \\ \hline
\end{tabularx}
\label{tab:use_cases}
\end{table}

In \ac{ODiS}, we focus on the power system layer's performance while the service level of the communication layer is fixed for the specific use case. The task for O-DMS and O-HEMS is to keep the power system layer in the normal state by using their self-* capabilities while disturbances appear both in the communication layer as well as the power system layer.


\section{Conclusion and Outlook}
\label{sec:conclusion}

The energy grid undergoes a dramatic change from a previously fully centralised to an increasingly distributed structure containing autonomous subsystems. The operation of this grid requires more observation and control mechanisms at all levels of the grid structure -- allowing for fast detection and reaction to unexpected developments, for instance. The Organic Distribution System (ODiS) initiative proposes to master the resulting issues of the control problem (such as robustness, scalability, efficiency, and limited autonomy) by means of introducing self-* capabilities as known from the Autonomic and Organic Computing domains. The main focus is on the four core self-* capabilities -- self-configuration, self-organisation, self-optimisation, and self-healing -- while methodologically establishing a framework that allows for later integration of further self-* mechanisms.

This article established a unified notion of terminology in the field and reviewed the state of the art in the field. We identified a gap in research towards an integrated cross-level solution, i.e. combining the distribution system functionality with those of home energy systems. Based on a system model for both levels, we identified parameters and decision freedom for autonomous or organic control modules operating in a distributed manner. This leads to a definition of the research scope of ODiS in the field.

Current and future work deals with the investigation and implementation of basic self-* functionality for ODiS. We presented the most urgent use cases and scenarios that allow the investigation of self-* mechanisms in simulation using the \textit{OpSim} framework. We further discussed the most promising evaluation setup and focus, which will be the scope of future investigations. Technically, we start with self-configuration and the corresponding self-awareness (i.e., state estimation) technology in simple use cases, before continuing with more sophisticated scenarios and closing the gap towards self-organisation, self-optimisation, and finally self-healing.

\up



\appendix

\section*{Bibliography}


\togglefalse{bbx:allnames}

\printbibliography[heading=none]


\end{document}

%% file: _acronyms.tex
\acrodef{ODiS}[ODiS]{\textbf{O}rganic \textbf{Di}stribution \textbf{S}ystem}
\acrodef{hems}[HEMS]{home energy management systems}
\acrodef{dms}[DMS]{distribution management systems}
\acrodef{ict}[ICT]{information and communication technology}
\acrodef{bsi}[BSI]{``Bundesamt f\"ur Sicherheit in der Informationstechnik''}
\acrodef{WP}[WP]{work packages}
\acrodef{MAS}[MAS]{multi-agent system}
\acrodef{EPS}[EPS]{electric power system}
\acrodef{CDE}[CDE]{controllable distributed energy unit}
\acrodef{DG}[DG]{distributed generators}
\acrodef{ML}[ML]{machine learning}


%% file: figures/InputOutput_ODMS_labels_new.tex
\begin{minipage}{\linewidth}
\centering
\begin{lpic}[]{figures/InputOutput_ODMS_new2(.77)}
  \lbl[.]{159,105;\large\textcolor[rgb]{0,0,0}{\fontsize{7pt}{6pt}\selectfont{$t:$ transformer, $s:$ switch, $g:$ generator}}}
 \lbl[l]{163,90;\large\textcolor[rgb]{0,0,0}{\begin{tabular}l{}\fontsize{8pt}{6pt}\selectfont{$\{P_t^k, Q_t^k, I_t^k, V_t^k, f^k\}$}\\\fontsize{8pt}{6pt}\selectfont{$\{P_g^k, Q_g^k, I_g^k, V_g^k\}, \delta_s^k$}
 \end{tabular}}} 
  \lbl[r]{154,90;\large\textcolor[rgb]{0,0,0}{\begin{tabular}r{}\scriptsize{$\{P_t^{{k,\star}}, Q_t^{{k,\star}}\}$}\\
  \scriptsize{$\{P_g^{{k,\star}}, Q_g^{{k,\star}}\}$, $\delta_s^{{k,\star}}$}\end{tabular}}}
 \lbl[l]{66,90;\large\textcolor[rgb]{0,0,0}{\begin{tabular}l{}\scriptsize{$\{P_t^{{k,\star}}, Q_t^{k,\star}\}, \{P_g^{{k,\star}}, Q_g^{{k,\star}}\}, \delta_s^{k,\star}$}\\
 \scriptsize{
  $\mathbb{I}_1^k, \mathbb{I}_2^k, ..., \mathbb{I}_n^k$}
 \end{tabular}}} 
    \lbl[r]{58.3,90;\large\textcolor[rgb]{0,0,0}{\begin{tabular}r{}\scriptsize{$\{P_t^k$, $Q_t^k$, $I_t^k$, $V_t^k$, $f^k\}$, $\delta_s^k\quad$}\\
    \scriptsize{$\{P_g^k, Q_g^k, I_g^k, V_g^k\}$,
 $\mathbb{O}_1^k, \mathbb{O}^k_2, ..., \mathbb{O}^k_n$}
 \end{tabular}}} 
 \lbl[l]{23,60;\large\textcolor[rgb]{0,0,0}{\begin{tabular}{l}\scriptsize{
  $\mathbb{I}_1^k$
 }\end{tabular}}} 
 \lbl[r]{16,60;\large\textcolor[rgb]{0,0,0}{
 \begin{tabular}{l}
 \scriptsize{
 $\mathbb{O}_1^k$}
 \end{tabular}}}
 \lbl[l]{64,60;\large\textcolor[rgb]{0,0,0}{\begin{tabular}{l}
 \scriptsize{
  $\mathbb{I}_i^k$
 }\end{tabular}}} 
 \lbl[r]{57,60;\large\textcolor[rgb]{0,0,0}{
 \begin{tabular}{l}
 \scriptsize{
 $\mathbb{O}^k_i$}
 \end{tabular}}}
 
 \lbl[l]{105,60;\large\textcolor[rgb]{0,0,0}{\begin{tabular}{l}
 \scriptsize{
  $\mathbb{I}_n^k$
 }\end{tabular}}} 
 \lbl[r]{97,60;\large\textcolor[rgb]{0,0,0}{
 \begin{tabular}{l}
 \scriptsize{
 $\mathbb{O}_n^k$}
 \end{tabular}}}
 
    \lbl[.]{175,15;\large\textcolor[rgb]{0,0,0}{\begin{tabular}{l}
  \fontsize{7pt}{6pt}\selectfont{$j:$ consumer/generator/prosumer}
  \end{tabular}
  }}
\lbl[l]{5,17.5;\large\textcolor[rgb]{0,0,0}{\begin{tabular}{l}
    \scriptsize{Input set of O-HEMS $i$:}\\
    \scriptsize{$\mathbb{I}_i^k = \{\underline{P}_i^{{k,\star}}\!, \overline{P}_t^{{k,\star}}\!, \underline{Q}_i^{{k,\star}}, \overline{Q}_i^{{k,\star}}\!,\underline{W}_i^{{k,\star}}\!, \overline{W}_i^{{k,\star}} \}$}\\  
    \scriptsize{Output set of O-HEMS $i$:}\\
    \scriptsize{$\mathbb{O}_i^k = \{\underline{\mathbf{P}}_i^{k}\!, \overline{\mathbf{P}}_i^{k}\!, \underline{\mathbf{Q}}_i^{k}, \overline{\mathbf{Q}}_i^{k}\!, \underline{\mathbf{W}}_i^{k}\!, \overline{\mathbf{W}}_i^{k}, P_i^k, Q_i^k, W_i
   ^k, I_i^k, V_i^k, f_i^k \}$}\\
    \scriptsize{with $i\in\{1,\ldots,n\}$ where $n\in \mathbb{N}$ is the number of O-HEMS}
  \end{tabular}
  }}
  \lbl[.]{210,56;\large\textcolor[rgb]{0,0,0}{
  \begin{tabular}{l}
\scriptsize{$P_{ij}^{{k,\star}}, $}\\\scriptsize{$Q_{ij}^{{k,\star}}, V_{ij}^{k,\star}$}
  \end{tabular}
  }}
 \lbl[.]{138.5,56.5;\large\textcolor[rgb]{0,0,0}{\begin{tabular}l{}
 \scriptsize{$P_{ij}^k, Q_{ij}^k, W_{ij}^k$}\\\scriptsize{ $I_{ij}^k, V_{ij}^k, f_i^k$}
 \end{tabular}
 }} 
\end{lpic}
\end{minipage}